\documentclass[aps,prd,preprintnumbers,nofootinbib,onecolumn]{revtex4}
\usepackage{bm}
\usepackage{latexsym}
\usepackage{dcolumn}
\usepackage{amsmath,amsfonts,amssymb}
\usepackage{graphicx,epsfig}
\usepackage{psfrag}
\usepackage{amsthm}

\def\be {\begin{equation}}
\def\ee {\end{equation}}
\def\bea {\begin{eqnarray}}
\def\eea {\end{eqnarray}}
\def\bc {\begin{center}}
\def\ec {\end{center}}
\def\bfg {\begin{figure}}
\def\efg {\end{figure}}
\def\bi {\begin{itemize}}
\def\ei {\end{itemize}}
\def\nn {\nonumber}

\def\la {\label}
\def\le {\left}
\def\ri {\right}

\def\fr {\frac}

\def\a  {\alpha}
\def\b  {\beta}

\def\D  {\Delta}

\def\beq{\begin{equation}}
\def\eeq{\end{equation}}
\def\br{\begin{eqnarray}}
\def\er{\end{eqnarray}}
\newcommand{\eel}[1] {\label{#1}\end{equation}}

\newcommand{\bdm}{\begin{displaymath}}
\newcommand{\edm}{\end{displaymath}}

\begin{document}
\title{Quantum Gravity Corrections and Entropy at the Planck time}


%

\author{Spyros Basilakos $^1$}\email{svasil@academyofathens.gr}
\author{Saurya Das $^2$}\email{ saurya.das@uleth.ca}
\author{Elias C. Vagenas $^1$}\email{evagenas@academyofathens.gr}

\affiliation{$^1$~Research Center for Astronomy and Applied Mathematics,\\
Academy of Athens, \\
Soranou Efessiou 4,
GR-11527, Athens, Greece}

\affiliation{$^2$~Theoretical Physics Group, Department of Physics and Astronomy,\\
University of Lethbridge, 4401 University Drive, Lethbridge,
Alberta -- T1K 3M4, Canada}

\begin{abstract}
We investigate the effects of Quantum Gravity on the Planck era of
the universe. In particular, using different versions of the
Generalized Uncertainty Principle and under specific conditions
we find that the main Planck quantities such as the Planck time,
length, mass and energy become larger by a factor of order
$10-10^{4}$ compared to those quantities which result from the
Heisenberg Uncertainty Principle. However, we prove that the
dimensionless entropy enclosed in the cosmological horizon at the Planck time
remains unchanged. These results, though preliminary, indicate that we should anticipate
modifications in the set-up of cosmology since changes in the Planck era will be inherited
even to the late universe through the framework of Quantum Gravity (or Quantum Field Theory)
which utilizes the Planck scale as a fundamental one. More importantly, these corrections
will not affect the entropic content of the universe at the Planck time
which is a crucial element for one of the basic principles of Quantum Gravity named Holographic Principle.

\end{abstract}

\maketitle

\section{Introduction}
\par
Gravity is a universal and fundamental force. Anything which has energy
creates gravity and is affected by it, although the smallness of
Newton's constant $G$ often means that the associated classical effects are
too weak to be measurable.
\par\noindent
An important prediction of various theories of quantum gravity
(such as String Theory) and black hole physics is
the existence of a minimum measurable length \cite{guppapers}.
The prediction is largely model-independent, and can be
understood as follows: the Heisenberg
Uncertainty Principle (HUP), $\Delta x \sim \hbar/\Delta p$, breaks down
for energies close to the Planck scale, when the corresponding
Schwarzschild radius is comparable to the Compton wavelength
(both being approximately equal to the Planck length).
Higher energies result in a further increase of
the Schwarzschild radius, resulting in $\Delta x \approx \ell_{Pl}^2\Delta p/\hbar$.
\par\noindent
At this point, it should be stressed that limits on the measurement of
spacetime distances as well as on the synchronization of clocks
were put in much earlier studies \cite{Mead:1964zz}. These limitations
showed up when Quantum Mechanics (QM) and General Relativity (GR) were put together
under simple arguments. It is more than obvious that in this context where one attempts
to reconcile the principles of QM with those of GR there are several and
even diverging paths to follow \cite{AmelinoCamelia:1994vs}.
In this framework, two of the authors (SD
and ECV) tracked a new path and showed that certain effects of Quantum Gravity are
universal, and can influence almost any system with a well-defined
Hamiltonian \cite{dv1}. Although the resultant quantum effects are
generically quite small, with current and future experiments,
bounds may be set on certain parameters relevant to quantum
gravity, and improved accuracies could even make them measurable
\cite{dv1,dv2}.
\par\noindent
One of the formulations, among those existing in the literature, of the {\it Generalized Uncertainty
Principle} (GUP) and which holds at all scales, is represented by
\cite{guppapers}
%
\be
\Delta x_i \Delta p_i \geq \fr{\hbar}{2} [ 1 + \beta
\le((\Delta p)^2 + <p>^2 \ri) + 2\beta \le( \Delta p_i^2 + <p_i>^2\ri) ]~,~i=1,2,3
\label{uncert1}
\ee
where $p^2 = \sum\limits_{j=1}^{3}p_{j}p_{j}$,
$\beta=\beta_0/(M_{Pl}c)^2=\beta_0\ell_{Pl}^2/\hbar^2$, $M_{Pl}=$
Planck mass, and $M_{Pl} c^2=$ Planck energy $\approx 1.2 \times
10^{19}~GeV$.
It is normally assumed that the dimensionless parameter
$\beta_0$ is of the order of unity.
However, this choice renders Quantum Gravity effects too small to be measurable.
On the other hand, if one does not impose the above condition {\it a
priori}, current experiments predict large upper bounds on it,
which are compatible with current observations, and may signal the
existence of a new length scale. Note that such an intermediate length
scale, $\ell_{inter} \equiv \sqrt{\beta_0} \ell_{Pl}$ cannot
exceed the electroweak length scale $\sim 10^{17}~\ell_{Pl}$ (as
otherwise it would have been observed). This implies $\b_0 \leq 10^{34}$.
Therefore, as stated above, Quantum Gravity effects influence all quantum Hamiltonians \cite{dv1}.
Moreover, some phenomenological implications of this interesting result  were presented in \cite{dv2}
\footnote{For a brief presentation of the results in \cite{dv1} and \cite{dv2} see \cite{dv3}.}.
\par
The recently proposed {\it Doubly Special Relativity} (or DSR)
theories on the other hand (which predict maximum observable momenta),
also suggest a similar modification of commutators \cite{AmelinoCamelia:2000mn,sm,cg}.
The commutators which are consistent with String Theory, Black Holes Physics,
DSR, {\it and} which ensure $[x_i,x_j]=0=[p_i,p_j]$ (via the Jacobi
identity) under specific assumptions lead to the following form \cite{asv1}
\bea
[x_i, p_j] = i \hbar\hspace{-0.5ex} \left[  \delta_{ij}\hspace{-0.5ex}
- \hspace{-0.5ex} \alpha\hspace{-0.5ex}  \le( p \delta_{ij} +
\frac{p_i p_j}{p} \ri)
+ \alpha^2 \hspace{-0.5ex}
\le( p^2 \delta_{ij}  + 3 p_{i} p_{j} \ri) \hspace{-0.5ex} \ri]
\label{comm01}
\eea
where
$\alpha = {\alpha_0}/{M_{Pl}c} = {\alpha_0 \ell_{Pl}}/{\hbar}$.
\par\noindent
Equation (\ref{comm01}) yields, in $1$-dimension, to ${\cal O}(\a^2)$
\be
\Delta x \D p \geq \frac{\hbar}{2}
\le[
1 - 2\a <p> + 4\a^2 <p^2>
\ri]
\label{uncert2}
\ee
where the dimensional constant $\alpha$ is related to $\beta$
that appears in equation (\ref{uncert1}) through dimensional analysis
with the expression $[\beta] = [\alpha^2]$. However, it should be
pointed out that it does not suffice to connect the two constants $\alpha$ and $\beta$
through a relation of the form $\beta \sim \alpha^2$ in order to reproduce
equation (\ref{uncert1}) from (\ref{uncert2}), or vice versa.
Equations (\ref{uncert1}) and (\ref{uncert2})
are quite different and, in particular, the most significant difference is that in equation (\ref{uncert1}) all terms
appear to be quadratic in momentum while in equation (\ref{uncert2}) there is a linear term in momentum.
Commutators and inequalities similar to (\ref{comm01}) and
(\ref{uncert2}) were proposed and derived respectively in
\cite{kmm,kempf,brau}. These in turn imply a minimum measurable
length {\it and} a maximum measurable momentum (to the best of our
knowledge, (\ref{comm01}) and (\ref{uncert2}) are the only forms
which imply both)
\bea
\D x &\geq& (\D x)_{min}  \approx \alpha_0\ell_{Pl} \la{dxmin} \\
\D p &\leq& (\D p)_{max} \approx \frac{M_{Pl}c}{\a_0}~. \la{dpmax}
\eea
\par\noindent
%
%
%
%
It is normally assumed as in the case of $\beta_0$
that the dimensionless parameter $\a_0$ is of the order of unity,
in which case the $\a$ dependent terms are important only when
energies (momenta) are comparable to the Planck energy (momentum),
and lengths are comparable to the Planck length.
However, if one does not impose this condition {\it a priori}, then using the fact that
all quantum Hamiltonians are affected by the Quantum Gravity corrections as was shown in \cite{dv1} and
applying this formalism to measure a single particle in a box, one deduces that all measurable
lengths have to be quantized in units of $\alpha_{0} \ell_{Pl}$ \cite{asv1}.
\par\noindent
In order to derive the energy-time uncertainty principle, we employ the equations
\bea
\Delta x &\sim& c \;\tau\nn\\
\Delta p &\sim& \frac{\Delta E}{c}\nn
\eea
where $\tau$ is a characteristic time of the system under study,
and it is straightforward to get
\be
\D x \D p \approx \D E \; \tau ~.
\label{energytime1}
\ee
Substituting equation (\ref{energytime1}) in the standard HUP, one gets the energy-time uncertainty principle
\be
\D E \; \tau \geq \frac{\hbar}{2} ~.
\label{energytime2}
\ee
It should be stressed that the characteristic time $\tau$ is usually selected
to be equal to the Planck time $t_{Pl}$ in the context of cosmology.
\par\noindent
The scope of the present work is to investigate in a
cosmological setup what corrections, if any, are assigned
to physical quantities such as the mass and energy of the
universe at the Planck time.
In particular, our present approach, regarding the
 Quantum Gravity Corrections at the Planck time, has been based on
a methodology that presented in the book of Coles and Lucchin \cite{coles}.
Simply, in our phenomenological formulation instead of using the standard HUP
for deriving the Planck time (as done in [11]), we let ourselves to utilize various versions of the GUP
and basically apply the methodology presented in the previously mentioned book.
\\
The significance of Planck time {\it per se} is due to the fact that it
is really a ``turning point''  because from the birth of the universe
till the Planck time Quantum Gravity corrections are significant
(classical General Relativity does not work at all) while after
that General Relativity seems to work properly.
%
%
%
\section{GUP and Entropy at Planck time}
%
In this section we investigate the effects of Quantum Gravity
on the Planck era of the universe.
By employing the different versions of GUP presented before,
we evaluate the modifications to several quantities that characterize the Planck era,
i.e. Planck time, length, mass, energy, density, effective number density and entropy.
This will enhance our understanding of the consequences of the Quantum Gravity
in the universe during the Planck epoch and afterwards.
\subsection{HUP and the standard Planck era}
\par\noindent
Without wanting to appear too pedagogical, we briefly present how
one can derive some physical quantities at the Planck epoch starting
from the HUP.
Following the methodology of \cite{coles} (see page 110), we first write the HUP in the form
\be
\Delta E \;\tau \simeq \hbar
\label{hup1}
\ee
and we adopt as characteristic time $\tau$ of the system under study the Planck time, i.e.
$t_{Pl}$, for which the quantum fluctuations exist on the scale of Planck length, i.e. $\ell_{Pl} = c t_{Pl}$.
In addition, the uncertainty in energy can be identified with the Planck energy and thus
$\Delta E = E_{Pl}=M_{Pl}c^{2}$. Thus, the HUP as given in equation (\ref{hup1}) is now written as
\be
M_{Pl} c^{2}\; t_{Pl} \simeq \hbar~.
\label{hup2}
\ee
Since the universe at Planck time can be seen as a system of radius $\ell_{Pl}$, the
Planck mass can be written as
\be
M_{Pl} \simeq \rho_{Pl} \ell_{Pl}^{3}
\label{mass2}
\ee
where by employing the first Friedmann equation the Planck density , on dimensional grounds, reads
\be
\rho_{Pl} \simeq \frac{1}{G t^{2}_{Pl}}~.
\label{density1}
\ee
Substituting equations (\ref{mass2}) and (\ref{density1}) in equation (\ref{hup2}), one gets
\be
\frac{1}{G t^{2}_{Pl}} \ell_{Pl}^{3} c^{2}\; t_{Pl} \simeq \hbar~
\ee
and therefore one can easily prove that the Planck time is
\be
t_{Pl}\simeq \sqrt{\frac{\hbar G}{c^{5}}}\simeq
10^{-43}sec \;.
\label{plancktime}
\ee
All the other parameters are defined in terms of the Planck time modulus some constants.
Indeed, the Planck length, density, mass, energy, temperature and
effective number density are given by the following expressions
\be \ell_{Pl}\simeq \sqrt{\frac{\hbar G}{c^{3}}}\simeq 1.7\times
10^{-33}cm \;,\;\;\; \rho_{Pl}\simeq \frac{1}{Gt^{2}_{Pl}} \simeq
\frac{c^{5}}{\hbar G^{2}} \simeq 4\times 10^{93}g\;cm^{-3}\;,
\label{plancklen} \ee

\be
M_{Pl}\simeq \rho_{Pl} l^{3}_{Pl}\simeq
\sqrt{\frac{\hbar c }{G}}\simeq 2.5\times 10^{-5}g \;,\;\;\;
E_{Pl}\simeq M_{Pl} c^{2} \simeq
\sqrt{\frac{\hbar c^{5} }{G}}\simeq 1.2\times 10^{19}GeV \;,
\label{planckener}
\ee

\be
T_{Pl}\simeq \frac{E_{Pl}}{k_{B}} \simeq
\sqrt{\frac{\hbar c^{5} }{G}} k^{-1}_{B} \simeq 1.4\times 10^{32}K \;,\;\;\;
n_{Pl}\simeq l^{-3}_{Pl} \simeq
\left( \frac{c^{3}}{G \hbar} \right)^{3/2} \simeq 10^{98} cm^{-3} \;.
\label{plancktep}
\ee
%
%
%
\subsection{GUP quadratic in $\Delta E$}
%
%
\par\noindent
The corresponding energy-time GUP of equation (\ref{uncert1})
\be
\Delta E \; {\tilde t}_{Pl}\geq \frac{\hbar}{2}
\left[1+\beta_0 \frac{{\tilde \ell}_{Pl}^{2}}{\hbar^{2}}\frac{(\Delta E)^2}{c^2}\right]
\label{gup2}
\ee
where we have kept only the first GUP-induced term of order \cal{O}$(\beta_0)$.
Note that the tilde denotes quantities with respect to the GUP.
As expected, for $\beta_{0}=0$ the GUP boils down to the standard form dictated by the
Heisenberg result ($\Delta E \;\tau \simeq \hbar$).

From the previously presented formalism (see section I),
the uncertainty in energy $\Delta E$ at the Planck time
is of order of the modified Planck energy, i.e.
${\tilde E}_{Pl}={\tilde M_{Pl}}^{}c^2$, where the
modified Planck mass lies inside the
universe's horizon of scale of the modified Planck length, i.e.
${\tilde \ell}_{Pl}$, and expands as
\be
{\tilde M}_{Pl}={\tilde \rho}_{Pl}^{} {\tilde \ell}_{Pl}^{3}~.
\label{mass1}
\ee
The modified Planck density can be easily derived
from the first Friedmann equation (or,
from the definition of the dynamical time scale) and
be given by
\be
{\tilde \rho}_{Pl} \simeq \frac{1}{G {\tilde t}_{Pl}^{2}}
\label{rho1}
\ee
modulus some constants.
Substituting equations (\ref{mass1}) and (\ref{rho1})
in equation (\ref{gup2}), one gets an equation
for the Planck time which now has
been affected by the Quantum Gravity corrections, namely
\bea
 \tilde{\rho}_{Pl}{\tilde \ell}_{Pl}^{3}c^{2} \tilde{t}_{Pl}
&\simeq&  \frac{\hbar}{2}
 \left[
 1+\beta_0 \frac{c^{2}\tilde{t}_{Pl}^{2}}
 {\hbar^{2} }{\tilde \rho}^{2}_{Pl}c^{8}\tilde{t}^{6}_{Pl}
 \right]\\
\frac{{\tilde \ell}_{Pl}^{3}}{G \tilde{t}^{2}_{Pl}}c^{2} \tilde{t}_{Pl} &\simeq&  \frac{\hbar}{2}
 \left[
 1+\beta_0 \frac{c^{2}\tilde{t}_{Pl}^{2}}
 {\hbar^{2}} \frac{c^{8}}{G^{2}\tilde{t}^{4}_{Pl}}\tilde{t}^{6}_{Pl}
 \right]\\
\frac{c^5}{G}\tilde{t}^{2}_{Pl} &\simeq&  \frac{\hbar}{2}
 \left[
 1+\beta_0 \frac{c^{10}}{\hbar^{2} G^{2}}\tilde{t}^{4}_{Pl}
 \right]~.
\eea
Therefore, after some simple algebra, one gets the following equation
\be
\frac{\beta_{0}}{2}\left(\frac{c^{5}}{\hbar G}\right)^{2}\tilde{t}^{4}_{Pl} -
\left(\frac{c^{5}}{\hbar G}\right)\tilde{t}^{2}_{Pl} +
\frac{1}{2}\simeq 0 \;.
\label{equation1}
\ee
It is easily seen that if we choose $\beta_{0}$ to be strictly equal to zero
then the current solution of the above equation reduces practically to that
of the standard Planck time (cf. (\ref{plancktime})).
On the other hand, $\forall \beta_{0} \in (0,1]$ equation
(\ref{equation1}) has two real solutions of the form
\bea
\tilde{t}_{Pl} &=& \sqrt{\frac{G \hbar}{c^{5}}} f_{\pm}(\beta_{0})\nn\\
&=& t_{Pl} f_{\pm}(\beta_{0})
\label{solution}
\eea
where
\be
f_{\pm}(\beta_{0})=\left[\frac{1 \pm \sqrt{1-\beta_{0}}}{\beta_{0}}\right]^{1/2} \;.
\ee
It is worth noting that for $\beta_0 = 1$ [or, equivalently, $f_{\pm}(1)=1$]
we find $\tilde{t}_{Pl}=t_{Pl}$, despite the fact that
we have started from a completely different Uncertainty Principle. This implies
that in this specific case, i.e. $\beta_0 = 1$, the GUP-induced effects in equation (\ref{solution}) cannot be observed.

\begin{figure}[ht]
\mbox{\epsfxsize=10cm \epsffile{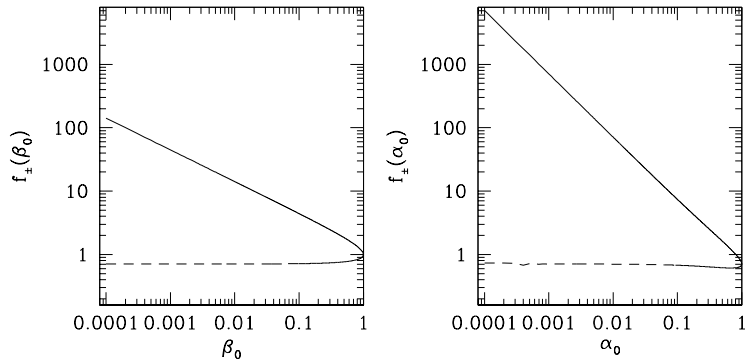}} \caption{
{\it Left Panel:} The predicted
Quantum Gravity corrections using the GUP of equation (\ref{gup2}).
{\it Right Panel:} The corresponding corrections utilizing
the GUP of equation (\ref{gup22}).
The solid and the dashed lines correspond to $f_{+}$ and
$f_{-}$, respectively.}
\end{figure}

Now we must first decide which is the important term
when $\beta_0$ takes values in the set $(0,1]$; is it $f_{-}(\beta_{0})$ or
$f_{+}(\beta_{0})$? Using some basic elements from calculus, one can prove that
the function $f_{-}(\beta_{0})$ is continuous and
increases strictly in the range of
$0<\beta_{0} \leq 1$
which implies that
$f_{-}(\beta_{0}) \in ({\rm lim}_{\beta_{0} \rightarrow 0} f_{-}(\beta_{0}),f_{-}(1)]$, where
${\rm lim}_{\beta_{0} \rightarrow 0}f_{-}(\beta_{0})=\sqrt{2}/2$.
Therefore, the
modified Planck time lies in the range
\be
\frac{\sqrt{2}}{2}t_{Pl} \leq \tilde{t}_{Pl} \leq t_{Pl}\;.
\ee
In the left panel of figure 1, we present
in a logarithmic scale the $f_{-}(\beta_{0})$ (dashed line)
as a function of $\beta_{0}$. Practically speaking,
the $f_{-}(\beta_{0})$ term has no effect on the Planck time.
On the other hand, following the latter analysis, we find that the
$f_{+}(\beta_{0})$ function (solid line)
decreases strictly in the range $0<\beta_{0} \leq 1$ which means that
$f_{+}(\beta_{0}) \in
[f_{+}(1),{\rm lim}_{\beta_{0} \rightarrow 0}f_{+}(\beta_{0}))$, where
${\rm lim}_{\beta_{0} \rightarrow 0}f_{+}(\beta_{0})=+\infty$.
It becomes evident that
as long as $\beta_{0} \ll 1$ [or, equivalently, $f_{+}(\beta_{0})\simeq (2/\beta_{0})^{1/2}$]
the current GUP can affect the Planck quantities via the function
$f_{+}(\beta_{0})$.
For example, in the case where $\beta_{0}={\cal O}(10^{-2}-10^{-4})$ we find
that $f_{+}\simeq {\tilde t}_{Pl}/t_{Pl}\simeq 10-10^{2}$.
\par\noindent
From a cosmological point of view, the ratio of the modified
Planck density, i.e. $\tilde{\rho}_{Pl}$, versus the measured dark energy, i.e. $\rho_{DE}\simeq
(1-\Omega_{m})\times 10^{-29}g\;cm^{-3}$ with $\Omega_{m}\simeq
0.27$ (for details see \cite{komatsu08}), is given as
\be
\frac{{\tilde \rho}_{Pl}}{\rho_{DE}}
\simeq \frac{10^{123}}{f^{2}_{+}} \simeq {\cal O} (10^{119}-10^{121}) ~.
\ee
One now is interested to investigate if and how
the main Planck quantities related to the Planck time are affected by the
above Quantum Gravity corrections. The corresponding relations are
\be
\label{quantities}
\frac{{\tilde \ell}_{Pl}}{\ell_{Pl}}=
\frac{{\tilde M}_{Pl}}{M_{Pl}}
=\frac{{\tilde E}_{Pl}}{E_{Pl}}=
\left(\frac{\rho_{Pl}} {{\tilde \rho}_{Pl}}\right)^{1/2}=
\left(\frac{{ n}_{Pl}}{\tilde n_{Pl}}\right)^{1/3}=
f_{+} \;.
\ee
Finally, it should be stressed that the dimensionless entropy
enclosed in the cosmological horizon of size $\tilde{\ell}_{Pl}$ now reads
\be
\tilde{\sigma}_{Pl} \simeq \frac{\tilde{\rho}_{P} c^{2} \tilde{\ell}^{3}_{Pl} }{k_{B} \tilde{T}_{Pl}}
\simeq \sigma_{Pl} \;.
\label{entropy1}
\ee
It is evident that the entropic content of the universe behind the
cosmological horizon at the Planck time is unaltered when Quantum Gravity
corrections are taken into account. Therefore, the information remains
unchanged: one ``particle" of Planck mass is ``stored"
in the Planck volume of the universe at the Planck time
behind the cosmological horizon of size $\tilde{\ell}_{Pl}$.
%
\subsection{GUP versus all terms of $\Delta E$}
%
%
%
\par\noindent
The corresponding energy-time GUP of equation (\ref{uncert2})
becomes
\be
\Delta E \; {\tilde t}_{Pl} \geq \frac{\hbar}{2}\left[1-2\alpha \frac{(\Delta
    E)}{c}+4\alpha^{2}\frac{(\Delta E)^{2}}{c^{2}}\right]~.
\label{gup22}
\ee
As it is anticipated, for $a=0$ (or, equivalently, $a_{0}=0$ since
$\alpha=\alpha_{0}\frac{c {\tilde t}_{Pl}}{\hbar}$) the GUP
boils down to the standard form dictated by HUP.
Evidently, performing the same methodology as before (see subsection B),
we obtain the following equation
\be
\label{new22}
2\alpha^{2}_{0} \left(\frac{c^5}{\hbar G}\right)^{2}\tilde{t}^{4}_{Pl}-
\frac{c^5}{\hbar G}(1+\alpha_{0})\tilde{t}^{2}_{Pl}+\frac{1}{2} \simeq
0 \;.
\ee
In deriving equation (\ref{new22}) we have substituted the
various terms as $\Delta E\simeq {\tilde E}_{Pl}$,
$\tau \simeq {\tilde t}_{Pl}$,
${\tilde \ell}_{Pl}  \simeq c {\tilde t}_{Pl}$,
and
$\alpha=\alpha_{0}\frac{c {\tilde t}_{Pl}}{\hbar}$.
In this framework, equation (\ref{new22}) has two real solutions
$\forall \alpha_{0} \in [-1/3,0)\cup (0,1]$. These are
\be
\tilde{t}_{Pl}=t_{Pl} f_{\pm}(\alpha_{0})
\label{solution22}
\ee
where
\be
f_{\pm}(\alpha_{0})=
\left[ \frac{(1+\alpha_{0})\pm
\sqrt{(1-\alpha_{0})(1+3\alpha_{0})}
}{4\alpha^{2}_{0}}
\right]^{1/2} \;.
\label{solution22}
\ee
Again it is routine to estimate the
limiting values of $f_{\pm}(\alpha_{0})$
$$f_{\pm}(1)=\frac{\sqrt{2}}{2}\;\;\;\;\;
f_{\pm}(-\frac{1}{3})=\sqrt{\frac{3}{2}}\;\;\;\;\;
{\rm lim}_{\alpha_{0} \rightarrow
  0}f_{-}(\alpha_{0})=\frac{\sqrt{2}}{2} \;\;.$$
Therefore, the function $f_{-}(\alpha_{0})$ does not play
a significant role (see the dashed line in the right panel
of figure 1), since the modified Planck time tends to the
usual value (${\tilde t}_{Pl}\approx t_{Pl}$).
On the contrary, if we consider the case of $f_{+}(\alpha_{0})$
(see the solid line in the right panel of figure 1), then
it becomes evident that for small values
of $\alpha_{0}$ the function $f_{+}(\alpha_{0}) \approx 1/2\alpha_{0}$
goes rapidly to infinity. As an example for
$\alpha_{0}={\cal O}(10^{-2}-10^{-4})$ we find
that $f_{+}\simeq {\tilde t}_{Pl}/t_{Pl}\simeq 10^{2}-10^{4}$.
Thus the ratio of the modified Planck density versus the measured dark energy now reads
\be
\frac{{\tilde \rho}_{Pl}}{\rho_{DE}} \sim {\cal O} (10^{115}-10^{119}) ~.
\ee
The main Planck quantities related to the Planck time are affected by the
above Quantum Gravity corrections exactly in the same way as shown in equation
(\ref{quantities}) employing the current form of $f_{+}(\alpha_0)$ defined in equation (\ref{solution22}).

\par\noindent
Furthermore, it is interesting to point out that the dimensionless entropy
enclosed in the cosmological horizon of size $\tilde{\ell}_{Pl}$ remains
unaltered $\tilde{\sigma}_{Pl}\simeq \sigma_{Pl}$.
This result is in accordance with the result derived in
previous subsection (see equation (\ref{entropy1})).
Therefore, the information in the Planck volume remains unchanged
even if one takes into account the Quantum Gravity effects,
i.e. one ``particle" of Planck mass is ``stored" in
the Planck volume of the universe at the Planck time
behind the cosmological horizon of size $\tilde{\ell}_{Pl}$.
\section{Conclusions}
In this work we have investigated analytically the
Quantum Gravity corrections at the Planck time by employing
a methodology that was introduced in the book of Coles and Lucchin \cite{coles}.
Specifically, in this work instead of using the standard HUP
for deriving the Planck time (as done in [11]), we let ourselves to utilize various versions of the GUP.
From our analysis, it becomes evident that the Planck quantities,
predicted by the Generalized Uncertainty
Principle GUP, extends nicely to those of the usual Heisenberg
Uncertainty Principle (HUP) and connects smoothly to them.
We also find that under of specific circumstances the modified Planck
quantities defined in the framework of the GUP
are larger by a factor of $f_{+}\sim (10-10^{4})$
with respect to those found using the standard HUP.

These results indicate that we anticipate modifications in the
framework of cosmology since changes in the Planck epoch will be
inherited to late universe through Quantum gravity (or Quantum
Field Theory). As an example, the calculation of the density
fluctuations at the epoch of  inflation sets an important limit on
the potential of inflation. Indeed, in the context of slow-roll
approximation, one can prove that the density fluctuations are of
the form $\delta_{H} \sim H^{2} / \dot{\phi}^2 = 3 H^{3} / V'$
where $\phi$ is the scalar field called inflaton, $H$ stands for
the Hubble parameter and $V(\phi)$ is the potential energy of the
scalar field. Assuming that $V \simeq m^{2}\phi^{2}$ where $m$ is
the inflaton mass, and $H \simeq \sqrt{V} / M_{Pl}$, we obtain
$\delta_{H} \simeq m \phi^{2} / M^{3}_{Pl}$. In order to achieve
inflation the scalar field has to satisfy the inequality $\phi
\geq M_{Pl}$ \cite{peacock}. Combining the above equations and
utilizing the observational value $\delta_{H}\approx 10^{-5}$, one
gets $m\leq 10^{-5} M_{Pl}$. Employing the Quantum Gravity
corrections, the latter condition becomes $\tilde{m} \leq 10^{-5}
f_{+}M_{Pl}$.

From the latter calculations it becomes evident that
the Quantum Gravity corrections affect directly the
main cosmological quantities (such as the inflaton mass) in the early
universe, via the $f_{+}=\tilde{t_{Pl}}/t_{Pl}$ factor. Due to the fact
that the $f_{+}$ factor is the consequence of the Quantum Gravity {\it per se}
(based on the GUP) implies that it can not be re-absorbed by a
redefinition of units.

From the observational point of view one can study the latter
corrections in the context of the primordial gravitational waves
which can be detected with very sensitive measurements of the
polarization of the CMB (see page 5 in \cite{Krauss:2010ty}). It
is interesting to mention here that the polarization of the CMB
will be one of the main scientific targets of the next generation
of the CMB data based on the Planck satellite. Therefore, if in
the near future the observers measure such an effect in the CMB
data then we may open an avenue in order to understand the
transition from the mainly-quantum gravitational regime to the
mainly-classical regime. We have already started to investigate
theoretically the above possibility and we are going to present
our results in a forthcoming paper.

Furthermore, it was also shown that the dimensionless entropy
enclosed in the cosmological horizon does not ``feel" the
Quantum Gravity corrections and thus the information
remains unaltered. Therefore, the entropic content of
the universe at the Planck time remains the same. This is quite important since
entropy is the cornerstone for one of the basic principles of Quantum Gravity
named Holographic Principle and for its incarnation known as AdS/CFT. \\

\par\noindent
{\bf Acknowledgments }\; The authors thank the anonymous referee
for useful comments and suggestions. SD was supported in part by
the Natural Sciences and Engineering Research Council of Canada
and by the Perimeter Institute for Theoretical Physics.


\end{document}